\documentclass[conference]{IEEEtran}
\IEEEoverridecommandlockouts

\usepackage{cite}
\usepackage{amsmath}
\usepackage{amssymb}
\usepackage{amsfonts}
\usepackage{multirow}
\usepackage{bm}
\usepackage[ruled,vlined,linesnumbered]{algorithm2e}
\usepackage{graphicx}
\usepackage{textcomp}
\usepackage{xcolor}
\usepackage[colorlinks=true,linkcolor=blue,urlcolor=black,citecolor=blue,bookmarksopen=true]{hyperref}
\usepackage{physics}
\usepackage{soul}
\usepackage[utf8]{inputenc}

\def\BibTeX{{\rm B\kern-.05em{\sc i\kern-.025em b}\kern-.08em
    T\kern-.1667em\lower.7ex\hbox{E}\kern-.125emX}}

\begin{document}

\title{Design of an entanglement purification protocol selection module\\

\author{\IEEEauthorblockN{1\textsuperscript{st} Yue Shi}
\IEEEauthorblockA{
\textit{Pacific Northwest National Lab}\\
Richland, WA \\
yoyoshi@uw.edu}
\and
\IEEEauthorblockN{2\textsuperscript{nd} Chenxu Liu}
\IEEEauthorblockA{
\textit{Pacific Northwest National Lab}\\
Richland, WA \\
chenxu.liu@pnnl.gov}
\and
\IEEEauthorblockN{3\textsuperscript{rd} Samuel Stein}
\IEEEauthorblockA{
\textit{Pacific Northwest National Lab}\\
Richland, WA \\
samuel.stein@pnnl.gov}
\and
\IEEEauthorblockN{4\textsuperscript{th} Meng Wang}
\IEEEauthorblockA{
\textit{The University of British Columbia}\\
Vancouver, BC, Canada \\
m.wang@ubc.ca}
\and
\IEEEauthorblockN{5\textsuperscript{th} Muqing Zheng}
\IEEEauthorblockA{
\textit{Pacific Northwest National Lab}\\
Richland, WA \\
muqing.zheng@pnnl.gov}
\and
\IEEEauthorblockN{6\textsuperscript{th} Ang Li}
\IEEEauthorblockA{
\textit{Pacific Northwest National Lab}\\
Richland, WA \\
ang.li@pnnl.gov}
}
}
\maketitle

\begin{abstract}

Entanglement purification protocols, designed to improve the fidelity of Bell states over quantum networks for inter-node communications, have attracted significant attention over the last few decades. These protocols have great potential to resolve a core challenge in quantum networking of generating high-fidelity Bell states. However, previous studies focused on the theoretical discussion with limited consideration of realistic errors. Studies of dynamically selecting the right purification protocol under various realistic errors that populate in practice have yet to be performed. In this work, we study the performance of various purification protocols under realistic errors by conducting density matrix simulations over a large suite of error models. Based on our findings of how specific error channels affect the performance of purification protocols, we propose a module that can be embedded in the quantum network. This module determines and selects the appropriate purification protocol, considering not only expected specifications from the network layer but also the capabilities of the physical layer. Finally, the performance of our proposed module is verified using two benchmark categories. Compared with the default approach and exhaustive search approach, we show a success rate approaching 90\% in identifying the optimal purification protocol for our target applications. 

\end{abstract}


\section{Introduction}

Advancements in quantum physics and engineering have catalyzed the development of a multitude of quantum computing platforms, from trapped ions to superconducting~\cite{Bluvstein:2024aa, Acharya:2023aa, PhysRevX.11.041058, Zhu:2023aa}. However, quantum computers encounter significant limitations in their noise-resilience, capacity, and scalability. Surmounting these errors and engineering a scalable quantum computer that can execute programs without succumbing to overwhelming levels of noise is of utmost importance. Quantum networking is a proposed approach to scale up systems by connecting multiple distributed quantum computers, reducing the demand on single QPU scalability and enabling the implementation of large-scale tasks \cite{A_quantum_network_stack_and_protocols_for_reliable_entanglement-based_networks,Quantum_Internet_protocol_stack_A_comprehensive_survey, Designing_a_quantum_network_protocol}.

For practical implementation of applications over quantum networks, especially for building reliable communication channels, entanglement is one of the most important resources. Entanglement pairs (EPs), in particular, are the most basic and indispensable building blocks of inter-node communication. However, in practice, EP quickly loses its information due to a multitude of noise sources, including but not limited to gate noise, thermal relaxation, and photon loss over optical networks. 

One way to mitigate this degradation and improve the fidelity of EPs is entanglement purification \cite{Bennett_purification,Deutsch_purification, Dur2007}. This technique trades off multiple EPs to produce a higher fidelity EP via local operation and classical communication (LOCC).  A multitude of purification protocols have been proposed and discussed. However, there are three remaining questions that prior purification studies have not been able to solve. \emph{Firstly,} The LOCC used for purifying EPs contains inevitable errors. However, previous studies of purification protocols are confined to discussions under the idealistic noiseless environment or have limited consideration of noise. \emph{Secondly,} Given the diversity of hardware imperfections, the optimal purification protocol for different types of hardware can be distinct. The discussion on the performance of purification protocols considering these variations is yet to be performed. \emph{Thirdly,} The practical design of a quantum network that leverages purification protocols to facilitate the quantum network computation is still in its nascency. We propose our work to partially address each of these challenges, generating a system-aware approach to model entanglement purification, with heuristic-based approaches to select protocols for actual hardware environments.

\begin{figure}
    \centering
    \includegraphics[width = 3.2 in]{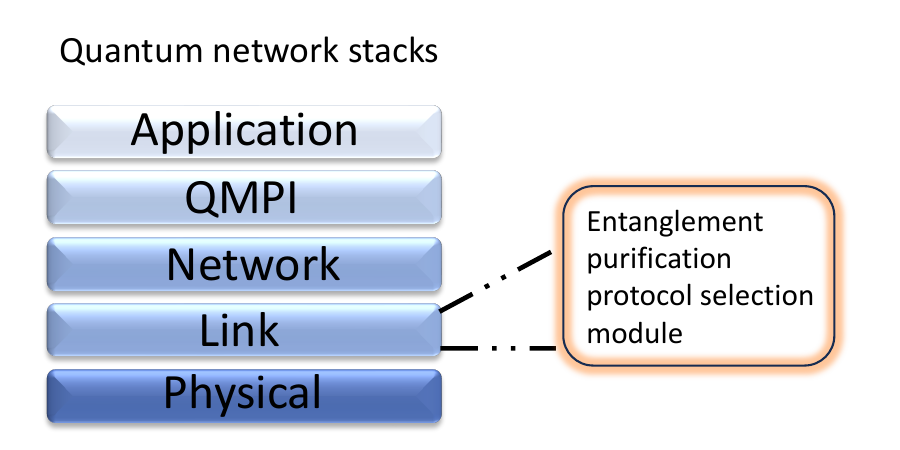}
    \caption{A quantum network with the Application, QMPI, Network, Link, and Physical layers. The proposed module embedded in the link layer is designed to select the appropriate entanglement purification protocol, considering not only the capabilities of the physical layer but also the expected specifications of the network layer in the quantum network.}
    \label{fig_1}
\end{figure}

In our work, we perform a comprehensive study on the performance of the existing EP purification protocols over density matrix simulation. According to the simulation results, we propose an entanglement purification selection module to dynamically select the appropriate protocol for the various hardware platforms. Our core contributions are summarised below: 

\begin{enumerate}

    \item We study state-of-the-art purification protocols considering a comprehensive suite of noise sources that affect the performance of the purification - from the size of "quantum memory" and rate of EPs generation to various errors on the buffer including depolarizing, thermal relaxation, and measurement errors.
    
    \item We observe the performance of various purification protocols under different noise settings via our density matrix simulator and shed light on the heuristics that go into selecting the purification protocol for different systems.

   \item We propose a module based on the aforementioned discovered heuristics that can be embedded in the link layer of a quantum network stack (shown in Fig. \ref{fig_1}) to select optimal purification protocols. 

\end{enumerate}

The rest of the paper is organized as follows. In the background section, we briefly introduce the involved purification protocols and various error channels. Then, we present the design of the auto-selection module in detail, including the parameters considered in the design, the comprehensive understanding of purification protocols under errors, and the principles guiding the design of our module. Subsequently, we present and discuss the simulation results from the density matrix simulation. Finally, we discuss the potential areas for further work. 

\section{Background}

\subsection{EP Purification protocols}

Various protocols have been proposed to purify lower purity entanglement resources, including bipartite entanglement, GHZ states, and cluster states \cite{Bennett_purification, Deutsch_purification, Dur2007, multipartite_entanglement_purification, PhysRevLett.95.220501,PhysRevA.57.R4075,SciRep.9.314(2019)}. In particular, protocols to purify EPs have received substantial interest. Existing protocols vary in a few ways: the number of pairs required, the sequence of local gates; and the resulting purification efficiency. In this subsection, we introduce three major protocols involved in the following. 

The density matrix of a raw EP is represented as ${\rho_{0}}$. The initial state with $n$ EPs is described by the density matrix ${\rho_{0}}\otimes{\rho_{1}}...\otimes{\rho_{n}}$. The fidelity of the imperfect EP to a perfect Bell state can be calculated by
\begin{equation}
    {F(\rho_{0})} = \langle\phi^+|\rho_{0}|\phi^+\rangle,
\end{equation}
where $\ket{\phi^+} = (\ket{00} + \ket{11})/\sqrt{2}$.

\subsubsection{\textbf{BBPSSW}}
The BBPSSW purification protocol, proposed by C. H. Bennett \textit{et al.} \cite{Bennett_purification}, is designed to purify the Werner states. The quantum circuit to implement the BBPSSW purification protocol in this work is shown in Fig.\ref{fig_protocol} (a). Once two EPs are prepared, bilateral CNOT gates are applied locally between the two copies, $U_{cnot}^{A_1 \rightarrow A_2}\otimes U_{cnot}^{B_2 \rightarrow B_1}$. Then, the second pair of qubits are measured against the computational \texttt{Z} basis. If the parity of the \texttt{ZZ} measurement is even, the fidelity of the target state increases. The increased fidelity, \( F'\), is described as follow:
\begin{equation}
F' = \frac{F^2 + [(1 - F)/3]^2}{F^2+2F(1-F)/3+5[(1-F)/3]^2}.
\end{equation}
Ideally, iteratively applying entanglement purification would keep improving the fidelity of the target EP. However, in the presence of noises, the potential improvement one can attain through purification is limited by system errors.

\subsubsection{\textbf{DEJMPS}}

In Ref~\cite{Deutsch_purification}, Deutsch \textit{et al.} proposed an advanced purification protocol, named DEJMPS. Unlike the BBPSSW protocol, the DEJMPS protocol does not require the initial states to be Werner states, making it more experimentally feasible. 

The quantum circuit to implement a DEJMPS purification protocol on devices is shown in Fig. \ref{fig_protocol}(b). After two EPs are initialized, \texttt{S} and \texttt{$\text{S}^\dagger$} are applied to flip the diagonal components of the initial density matrix $\rho_{0}$ to the Bell-diagonal form. Then, bilateral \texttt{CX} gates are executed on qubit pairs ($q_{0}$, $q_{2}$) and ($q_{1}$, $q_{3}$), respectively. Finally, the second EP, ($q_{2}$, $q_{3}$) is measured in \texttt{ZZ} basis. If the measurement results coincide between two qubits, the fidelity of the target state increases. 

Suppose the initial state of the raw EP is 
\begin{equation}
    \rho_0 = A \dyad{\phi^+} + B \dyad{\phi^-} + C \dyad{\psi^+} + D \dyad{\psi^-},
\end{equation}
where $\ket{\phi^{\pm}}$ and $\ket{\phi^\pm}$ are four Bell states, $A$, $B$, $C$, $D$ are the corresponding coefficients. The fidelity of the initial state to $\ket{\phi^+}$ state is $F = A$. After a successful purification operation, the fidelity to $\ket{\phi^+}$ state can be improved to 
\begin{equation}
    F' = (A^2 + B^2)/N,
\end{equation}
where $N = (A + B)^2 + (C + D)^2$ is the success probability.

\subsubsection{\textbf{EXPEDIENT}}

\begin{figure}[ht]
    \centering
    \includegraphics[width = 3.4 in]{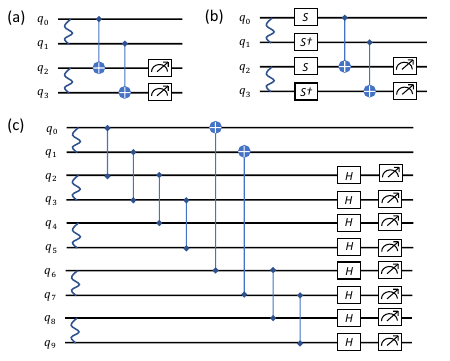}
    \caption{The quantum circuits to conduct one round of \textbf{(a)} BBPSSW \textbf{(b)} DEJMPS and \textbf{(c)} EXPEDIENT entanglement purification protocol on a device.}
    \label{fig_protocol}
\end{figure}

The EXPEDIENT purification protocol, proposed by Nickerson \textit{et al.},\cite{expedient_purification} requires five copies of EP to purify towards a single EP. In the previous study, EXPEDIENT outperforms the other protocols in terms of fidelity improvement under limited errors\cite{expedient_purification}. Fig. \ref{fig_protocol}(c) shows the quantum circuit to implement one round of the EXPEDIENT protocol. Compared with other protocols, it requires significantly more EPs, and a deeper circuit for each round of purification. Therefore, although it outperforms in theoretical studies, its performance on realistic devices may not necessarily be better than that of other protocols due to its increased demand on topology, gate fidelity, and number of EPs.

\begin{figure*}[ht]
    \centering
    \includegraphics[width = 6.2 in]{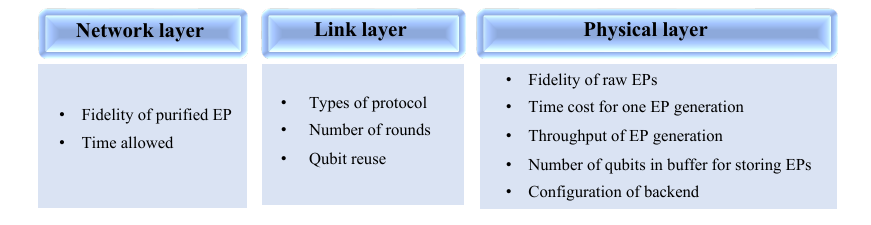}
    \caption{The diagram illustrates the parameter spaces for the purification protocol selection, deriving from three adjacent stacks in quantum networks.}
    \label{fig2}
\end{figure*}
\subsection{Error models}

Quantum computing suffers noise from various sources, like imprecise control, decay and decoherence, etc. In this subsection, we introduce the error models considered in our simulation.

Quantum noise is modeled by quantum channels $\mathcal{E}$, which is represented by a set of Kraus operators ($\{ E_{k}\}$). The quantum channel transforms the state of a quantum system according to 
\begin{equation}
    \rho = \mathcal{E}(\rho_0) = \sum_{k} E_k \rho_{0} E_k^\dagger,
\end{equation}
where $\rho_0$ are initial state, $\rho$ is the final state.
As the transformation preserves the trace of the density matrix, the Kraus operators of a quantum channel should satisfy
\begin{equation}
    \sum_k E_k E_k^\dagger = I,
\end{equation}
where $I$ is the identity operator~\cite{Nielsen_Chuang_2010}.

Next, we introduce the Kraus operators of different types of errors involved in this work below.

\subsubsection{\textbf{Depolarizing}}
The depolarizing error refers to a type of error that randomly changes the state of the qubits towards a mixed state. The depolarizing channel for a single qubit with an error rate $p$ is described by the following Kraus operators:
\begin{equation}
\begin{split}
E_0 = \sqrt{1 - \frac{3p}{4}} I, \quad 
E_1 = \sqrt{\frac{p}{4}} X, \\
E_2 = \sqrt{\frac{p}{4}} Y, \quad 
E_3 = \sqrt{\frac{p}{4}} Z,
\end{split}
\end{equation}
where $\{I, X, Y, Z\}$ are Pauli matrices. $E_{0}$ describes the qubit remaining in the origin state. $E_{1}$, $E_{2}$ and $E_{3}$ describes the qubit undergoing a Pauli $X$, $Y$ or $Z$ error.

\subsubsection{\textbf{Measurement error}}

Measurement error refers to incorrectly assigning an observation based on the process of measuring a quantum system. The measurement error channel for measuring the ground state $|0\rangle$, with the error probability $p_{01}$ of measuring the state as $|1\rangle$, is described by the following two Kraus operators:
\begin{equation}
E_0 = \begin{pmatrix}
1 & 0 \\
0 & \sqrt{1 - p_{01}}
\end{pmatrix}I, \quad
E_1 = \begin{pmatrix}
0 & \sqrt{p_{01}} \\
0 & 0
\end{pmatrix}Z.
\end{equation}
$E_0$ describes the possibility of measuring correct state $|0\rangle$, $E_1$ describes the possibility that the measurement yields the incorrect state $|1\rangle$. The measurement error for measuring a quantum state $|1\rangle$, with error possibility $p_{10}$ of measuring the state as $|0\rangle$ can be described using similar operators.

\subsubsection{\textbf{Reset error}}
Reset error appears when a state of qubit is reset to a specific state, commonly $|0\rangle$ or $|1\rangle$. The reset error channel for resetting a qubit with state $|u\rangle$ to ground state $|0\rangle$ but with probability $p$ of resetting to $|1\rangle$ is described by following Kraus operators: 
\begin{equation}
E_0 = \sqrt{1 - p} |0\rangle \langle u|, \quad
E_1 = \sqrt{p} |0\rangle \langle u|.
\end{equation}
$E_{0}$ describes the qubit being correctly reset to $|0\rangle$. $E_{1}$ describes the qubit being reset to $|1\rangle$.

\subsubsection{\textbf{Amplitude damping}}
Amplitude damping refers to the process where a quantum state loses excitation, which is associated with the lifetime $T_{1}$ of qubits. The amplitude-damping channel for a single qubit is described by the following Kraus operators:
\begin{equation}
E_0 = \begin{pmatrix}
1 & 0 \\
0 & \sqrt{1 - \lambda}
\end{pmatrix}, \quad
E_1 = \begin{pmatrix}
0 & \sqrt{\lambda} \\
0 & 0
\end{pmatrix}.
\label{eq_amp_damping}
\end{equation}
$E_0$ describes the qubit remaining in its current state without decay. $E_1$ describes the qubit decaying from the excited state to the ground state. $\lambda$ is the amplitude-damping rate, which is defined by:
\begin{equation}
\lambda = 1-\exp(\frac{-t}{T_{1}}).
\label{amplitude-damping}
\end{equation}
$t$ is the time interval for amplitude damping.

\subsubsection{\textbf{Phase damping}}
Phase damping refers to a process where a quantum state loses phase information without loss of energy, which is associated with both lifetime, $T_{1}$, and coherence time, $T_{2}$. The phase damping channel for a single qubit is described by the following two Kraus operators:
\begin{equation}
E_0 = \sqrt{1 - \frac{\lambda}{2}} I, \quad E_1 = \sqrt{\frac{\lambda}{2}} Z.
\end{equation}
$E_{0}$ describes non-phase damping occurring on the qubit. $E_{1}$ described the phase damping occurrence on the qubit. $\lambda$ is the phase-damping rate, which is defined as:
\begin{equation}
\begin{split}
\lambda = 1-\exp(\frac{-t}{T_{\phi}}), \\
T_{\phi} = \frac{T_{1}T_{2}}{2T_{1}-T_{2}}.  
\end{split}
\label{eq_phase}
\end{equation}
We include the error from quantum state idling by applying \texttt{Delay} gates on qubits in our simulation. The idling of qubits induces the thermal relaxation of the quantum state, leading to energy loss (amplitude damping) and loss of quantum coherence (phase damping). The group of Kraus operators to describe the thermal relaxation channel caused by idling is a sum of Kraus operators from amplitude damping and phase damping channels.

\section{Design of module}

In this section, we introduce the design of our purification protocol selection module. Our design takes into account multiple hardware properties including physical buffer size, coherence time, gate fidelity, and EP generation rate. Furthermore, we consider the specifications of the network, such as the expected fidelity of EPs and the maximum allowable time to obtain purified EPs. Based on these constraints, it determines an appropriate purification protocol to perform entanglement distillation with, whilst attempting to maximize the fidelity of EPs. 

In this section, we begin by introducing the parameter regimes considered in the design of this module. Following this, we present the potential fidelity improvements over a multitude of entangled pair fidelity and system errors. Finally, the design principles and heuristics that guide the development of this module are established. Here, we consider only bipartite shared entanglements in the form of EPs. The developed heuristic approach in this work can be extended and applied to more general cases where more than two parties are involved. We assume there is a buffer containing pairs of qubits between connected quantum devices to store the generated EPs, resulting in our purification protocols being executed in serial.

\subsection{Design parameters}

The performance of a purification protocol is significantly affected by a multitude of parameters. Here, we classify those parameters into two categories: the capability of the physical layer and the specifications expected from the network layer. These two categories of parameters are shown in two blocks for the network and physical layer in Fig. \ref{fig2}.

EPs generation is characterized using \texttt{$n$} and \texttt{$\tau$}. \texttt{$n$} denotes the throughput of one EP generation cycle. \texttt{$\tau$} denotes the average generation time, which is defined as $\tau = 1/r$ ($r$ is the generation rate). For the purification process that needs EPs from more than one generation, the stored EPs have to idle a time of \texttt{$\tau$}, which applies a degradation on EPs.

For the buffer connecting different quantum nodes, we will only consider the size of the buffer, \texttt{num\_qubits}, which represents the number of qubits available to store EPs for purification. We do not consider the architecture and the write-in and read-out working mechanisms of a buffer.

\texttt{Config} describes the configuration of the quantum buffer, specifying the physical properties, supported gate operations, connectivity, etc. For example, if the buffer is modeled based on one of the devices from IBMQ, the configuration is a set of experimental calibrated parameters, including coherence times  \texttt{T1/T2}, 
readout length \texttt{readout\_length}, measurement errors \texttt{prob\_meas0\_prep1} and \texttt{prob\_meas1\_prep0}, and gate times \texttt{gate\_length} and gate error rate \texttt{gate\_err}.

Besides the various errors in the physical layer, the specifications expected by the network layer are also crucial in determining an appropriate purification protocol. \texttt{F\textsubscript{out}} presents the minimum acceptable fidelity of the outcome EPs. \texttt{T\textsubscript{qos}} specifies the allotted time to obtain the purified EPs.

As shown in the ``Link Layer'' of Fig. \ref{fig2}, the implementation of the purification protocol introduces three tunable features. The first feature is the type of protocol. More advanced protocols in theory do not necessarily perform better on all devices. With realistic errors, the optimal protocol for a specific device varies and needs to be determined by a combination of numerical simulation and heuristics. The second parameter is the number of purification rounds, $n$. Under ideal conditions, all gates and operations are noise-free, and each round of purification will iteratively improve EPs fidelity. However, in practice, more numbers of purification rounds could, on the contrary, reduce the fidelity of EPs. The third feature is enabling qubit reuse. The nested implementation of the purification protocol is expected to yield EPs with a high fidelity. However, the number of qubits needed for the nested protocol exponentially increases as the number of purification rounds increases, which is constrained by the capability of the hardware. For example, the implementation of a two-round nested DEJMPS protocol requires four EPs prepared in the buffer. With only a six-qubit available in the buffer, a compromise way is enabling the reuse of some of the qubits. The EP purified by the first two EPs is stored in the buffer. Then the second two qubits are reset to ground states, while another EP is generated. This EP and the last EP are used for another purification sequence, yielding the second one-round purified EPs. Finally, two purified EPs are used for the second round of purification. The purification protocol, which scales exponentially in qubit demand, can be conducted on hardware with a limited number of qubits by enabling qubits reuse.

\begin{figure*}
    \centering
    \includegraphics[width = 5.5 in]{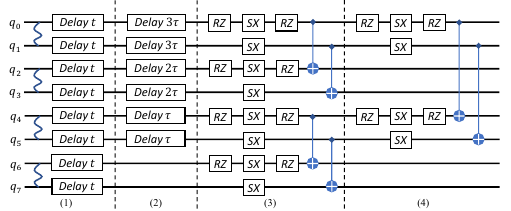}
    \caption{The quantum circuit used to simulate the nested two rounds of DEJMPS purification protocol with input fidelity, $F_\text{in}$, and waiting time, $\tau$, in our simulation. In the first region, four EPs are modeled using \texttt{RZ}, \texttt{SX}, and \texttt{CNOT} gates on each of two adjacent qubits. \texttt{Delay} is used to tune the fidelity of the raw EPs. In the second region, \texttt{Delay} with $n\times\tau$ ($n=1,2,3$) simulate the waiting period of the first coming EPs for the upcoming EPs. In the third and fourth regions, two rounds of DEJMPS purification are implemented using a set of basis gates. Finally, qubits $q_{2}$ to $q_{7}$ are measured and selected, $q_{0}$ to $q_{1}$ is yielded with higher fidelity when measured results coincide in the measurements on every two adjacent qubits.}
    \label{fig_qc}
\end{figure*}

Figure \ref{fig_qc} shows the quantum circuit to implement two rounds of nested DEJMPS protocol. There are eight qubits in this experiment, which forms four EPs. In the quantum communication application, we assume $q_{0}$, $q_{2}$, $q_{4}$, and $q_{6}$ belong to one end, and the rest of the qubits are held by another end. The gate operations are broken down into basis gates of IBM devices, \texttt{RZ}, \texttt{CNOT}, \texttt{X}, \texttt{SX}, \texttt{Delay} and \texttt{Reset}. Here, EP generation is modeled using \texttt{RZ}, \texttt{SX}, and \texttt{CNOT} gates. The \texttt{Delay} gates with duration $t$ are used to tune the fidelity of raw EP, mimicking delays in EP transportation and verifying fidelity improvement with different inputs. Here, we assume the EP generation time $\tau$ is 50 ns and the throughput of the generation $n$ is 1. Therefore, in the purification process, the first EP needs to wait 50 ns to get the second EP ready, and so on. The extra \texttt{Delay} simulates any excess idling of qubits while waiting for pending EPs. Following the implementation of two rounds of DEJMPS protocol, $q_{2}$ to $q_{7}$ are measured in \texttt{ZZ} basis. If the measured results coincide between two qubits of each EP, $q_{0}$ and $q_{1}$ obtain a relative increase in fidelity compared to the initial fidelity. 

\subsection{Effects of realistic errors}

To understand how realistic parameters discussed above affect the performance of purification protocols, we conducted a comprehensive study of various purification protocols under noise settings using a density matrix simulator \cite{li2020density}. The value of the input fidelity, $F_\text{in}$, and the value of each type of error, $\tau$, is swept. At each $F_\text{in}$ and $\tau$, purification protocols yield the fidelity improvement, $\Delta F$ ($\Delta F=F_\text{out}-F_\text{in}$). The maximum fidelity improvement, denoted as \texttt{$\Delta F_\text{max}$}, is selected among the fidelity improvements of all protocols and shown in different colors in the phase diagrams in Fig. \ref{sweeping}. For each subplot, only one error is included and denoted on the y-axis. Other errors are disabled for that simulation to prevent combinatorial explosion over noise model settings. 

\subsubsection{$\tau_\text{depolarizing}$}
$\tau_\text{depolarizing}$ denotes the error rate of local operations in the purification process that solely stem from quantum state depolarizing. We studied how depolarizing error solely affects the performance of protocols by setting the operation time of all gates involved to zero, ensuring that no thermal relaxation appears in the process. The error probability of local \texttt{CX} gates, $p$, are preset by overwriting in the backend configuration file. Under the depolarizing error, the off-diagonal elements in the density matrix of the initial state are reduced by a factor of $1-p$, and the diagonal elements are moved towards a uniform distribution.

As shown in Fig. \ref{sweeping}(a), one round of EXPEDIENT protocol outperforms others while the raw EPs have high fidelity. As the fidelity of raw EPs decreases, the DEJMPS protocol outperforms the other protocols. One round of the DEJMPS protocol yields the best fidelity improvement when the depolarizing error is small, while two rounds are more effective when the depolarizing error is close to 0.01. That is consistent with our understanding that more rounds of purification are better to run on less noisy hardware.

\subsubsection{$\tau_\text{amplitude\_damping}$} $\tau_\text{amplitude\_damping}$ denotes the error rate of local operations that solely stem from the amplitude damping of quantum states. The amplitude damping channel is modeled by Kraus operators shown in equation \ref{eq_amp_damping}.
The amplitude damping channel, not just affect the probabilities of the quantum state (diagonal element of the density matrix), but also the coherence in qubits (off-diagonal elements of the density matrix), leading to a significant impact on the fidelity of a quantum state.

To study how amplitude damping error solely affects the performance of protocols, we set the lifetime \texttt{T\textsubscript{1}} of each qubit to $5\times10^{-7}$ s, coherence time \texttt{T\textsubscript{2}} of each qubit to two times of the value of \texttt{T\textsubscript{1}}. The operation length of local \texttt{CX} gates is swept from $1\times 10^{-5}$ to $1\times10^{-1}$ s. The amplitude damping error rate, $\tau_\text{amplitude\_damping}$, is estimated using equation \ref{amplitude-damping} from \texttt{T\textsubscript{1}} and \texttt{gate\_length}. The fidelity of raw EP, $F_\text{in}$, spans from 0.83 to 1, captured by adjusting the delay time ($t$) from 500 to 0 ns in the EP preparation circuit (see Fig.~\ref{fig_qc}).

The simulation result is shown in Fig. \ref{sweeping}(b). We found the maximum improvement of fidelity $\Delta F_\text{max}$ is significantly affected by the $F_\text{in}$ and $\tau_\text{amplitude\_damping}$. For the upper right region of the subplot (above the straight line connecting $F_\text{in}\approx0.84$ and $\tau_\text{amplitude\_damping}=0.4$, and $F_\text{in}\approx0.94$ and $\tau_\text{amplitude\_damping}=0.2$), $\Delta F_\text{max}$ is below zero indicating that no purification protocol increases the fidelity of EPs with such preset parameters. Surprisingly, the best-performing protocol in the regime of small error is not the EXPEDIENT protocol but multiple rounds of nested DEJMPS protocol. The required amounts of EPs and the significant circuit depth of the EXPEDIENT protocol challenge efficiency. 

\subsubsection{$\tau_\text{phase\_damping}$}
$\tau_\text{phase\_damping}$ denotes the error rate of local operations in the purification process that solely stem from the phase damping.
The phase-damping channel affects the off-diagonal elements of the density matrix. It describes the loss of phase information of qubits without energy loss. Thus, phase-damping errors do not affect the diagonal elements of the density matrix or the probabilities of the computational \texttt{Z} basis states. 

We study the sole effect of phase damping on purification protocols by setting the \texttt{T\textsubscript{1}} to $1\times 10^{-3}$ s and \texttt{T\textsubscript{2}} to $1\times 10^{-6}$ s (\texttt{T\textsubscript{1}} $\gg$ \texttt{T\textsubscript{2}}), in which the thermal relaxation almost all comes from phase damping. The gate length of all local \texttt{CX} gates are from $1\times10^{-8}$ to $1\times10^{-6}$ s, which are typical gate times on IBMQ devices. The error rates, $\tau_\text{phase\_damping}$, are calculated from \texttt{T\textsubscript{1}}, \texttt{T\textsubscript{2}} and \texttt{gate\_length} using eq. \ref{eq_phase}. In our simulation, as shown in Fig. \ref{sweeping}(c), we found the choice of the best purification protocol under only phase damping error is independent of the value of the phase damping error, and only impacted by the input fidelity, $F_\text{in}$. This is in agreement with the performance of phase damping error on the density matrix. For $F_\text{in} > 0.58$, one round of the EXPEDIENT protocol shows the largest fidelity improvement; For $F_\text{in} < 0.58$, two rounds of the DEJMPS protocol show the presents better performance.

Our study suggests that for a hardware platform with phase damping error orders of magnitude larger than other noises, the selection of the EPs purification protocol mainly depends on the fidelity of raw EPs rather than the error rate.

\subsubsection{$\tau_\text{idling}$}

\begin{figure*}[ht]
    \centering
    \includegraphics[width = 6.4 in]{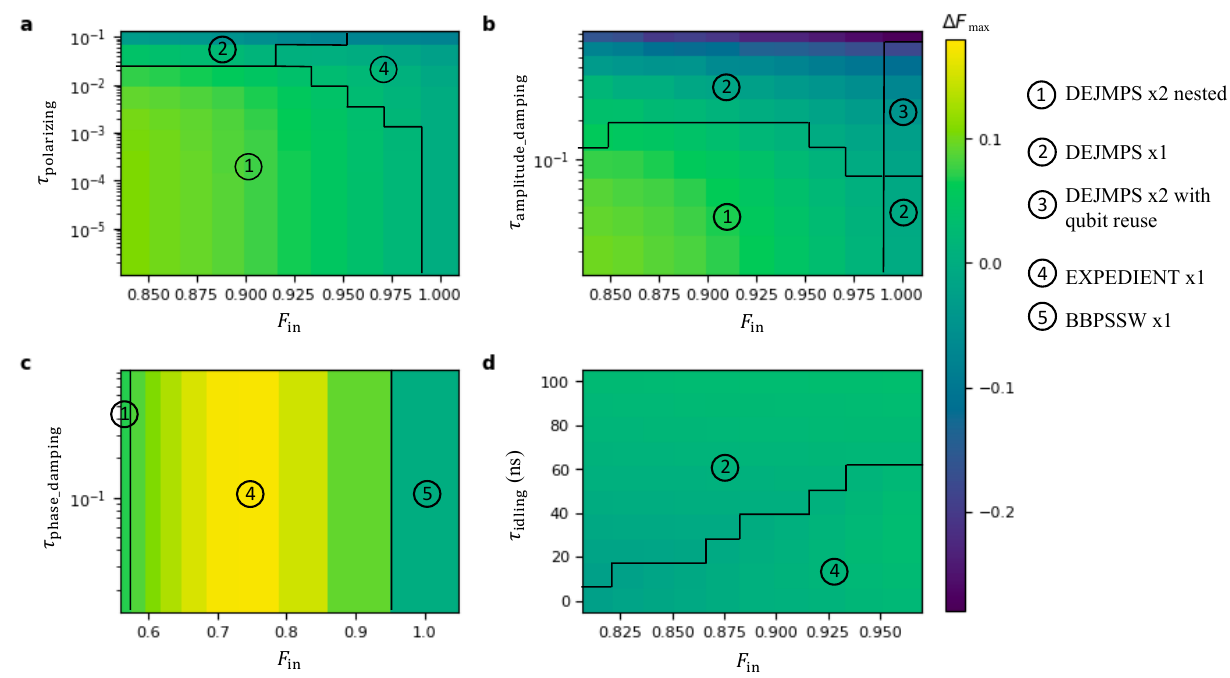}
    \caption{The largest fidelity improvement, $\Delta F_\text{max}$, over outputs of all purification protocol as the input fidelity, $F_\text{in}$, and one type of error (a) $\tau_\text{depolarizing}$,
 (b) $\tau_\text{amplitude\_damping}$, (c) $\tau_\text{phase\_damping}$, and (d) $\tau_\text{idling}$ swept. The numbers in the figures denote the corresponding protocol.}
    \label{sweeping}
\end{figure*}

We finally studied how the EP generation capacity affects the performance of the purification protocol in practice. $\tau_\text{idling}$ denotes the averaged time in nanoseconds for generating one EP, $\tau_\text{idling}=n/\tau$ ($n$ is the throughput of the EPs generation, $\tau$ is the time for each generation). When the throughput of EP generation is smaller than the number of EPs needed for the purification, the first generated EPs idle waiting for upcoming EPs, leading to the thermal relaxation of states. This thermal relaxation combines phase damping and amplitude damping, affecting both the off-diagonal and diagonal elements in the density matrix of state. The simulation result is shown in Fig. \ref{sweeping}. The one round of EXPEDIENT protocol outperforms with large input fidelity and small $\tau_\text{idling}$, while one round of DEJMPS protocol outperforms in the opposite regime.

Our study suggests that when applying purification protocol, extra attention is required when handling the EP generation rate. For a hardware setup where the generation time is non-negligible in the purification process, if the errors caused by EPs idling are significantly larger than other noise sources, the EXPEDIENT protocol is the best only for a higher input fidelity and shorter generation time.

\subsection{Design principles}
In this subsection, we introduce the design principles and the pseudo-code of our proposed module (shown in Algorithm. 1) as follows. The design principles include two steps of protocol pruning and one step of sorting. $P$ is the group of protocols that contains three types of protocols with different numbers of rounds and implementation methods.

a) Trim down $P$ according to the last reported noise profile of the device. The first pruning process of the protocols from group $P$ is determined by the capacity of the buffer and the compute device. For example, if the number of qubits exceeds the size of the buffer, or if the runtime needed for a protocol exceeds the allotted time, the protocol does not apply to this setup and is pruned.

b) The second pruning of protocols is based on the comprehensive understanding of each purification protocol under various errors presented in the phase diagrams shown in the last subsection. In this step, firstly, the error rate of each type of error, $e_{j}$ is estimated using the method introduced in the previous context and averaged over qubit channels involved. As shown in the pseudo-code, if any error is larger than $v_{1}$ ($v_{1} = 1\times 10^{-4}$), the protocol with more than two rounds is pruned from $P$. When the hardware is very noisy, multi-round purification is simply impossible, as rounds increase, noise contribution is greater than the amount of noise being purified. Meanwhile, if all types of errors are smaller than $v_{2}$ ($v_{2} = 1\times10^{-4}$), the purification implementations based on the BBPSSW protocol are pruned. The BBPSSW protocol, with fewer operations involved in the process, presents benefits for noisy devices but has the opposite for low-noise devices. As we know that the EXPEDIENT protocols do not favor the hardware with significant errors due to the depth of operations, our design gives one more boundary, $v_{3}$. For setup with $e_{j} > v_{3}$, $\exists j \in J$ ($v_{3} = 0.01$), the EXPEDIENT protocols are pruned from the group. The protocols group $P'$ is returned after these two pruning steps.

c) Finally, $P'$ is sorted based on our generated heuristics from the simulation. If $F_\text{in}$ already is approximated to meet the requirement, one can skip the EP purification process. Otherwise, the left protocols in $P'$ are sorted depending on the fidelity of raw EPs. Then, the first protocol in the sorted group is used. Within the allotted time, that is $T_\text{remain} > 0$, the purification protocol is continuously selected based on this approach and applied over the EPs to maximize fidelity while maintaining efficiency. Finally, the selected protocol list $P_\text{out}$ and purified EPs are delivered.

\begin{figure}
    \centering
    \includegraphics[width = 3.2 in]{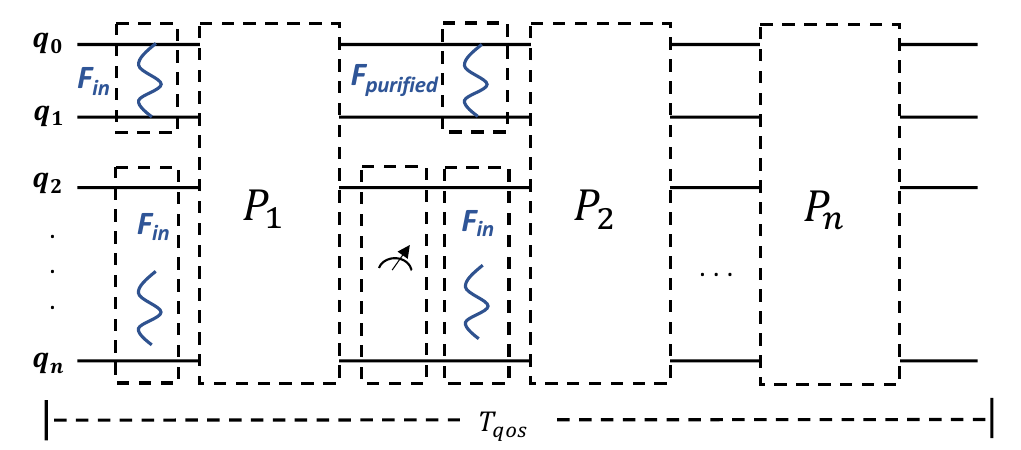}
    \caption{A diagram illustrates the repeated implementation of purification within allotted time \texttt{T\textsubscript{qos}}. In the first round of protocol selection, Protocol $P_{1}$ is selected and implemented. The qubits started from $q_{2}$ are measured and selected, and the first pair EP becomes $F_\text{purified}$. With the remaining time, further rounds of purification are sequentially implemented.}
    \label{fig_seq}
\end{figure}

Figure \ref{fig_seq} demonstrates the dynamic purification protocol selection and implementation process within the allotted time. The first protocol $P_{1}$ is selected and applied over input EPs with identical fidelity $F_\text{in}$, After that, $q_{2}$ to $q_{n}$ are measured. The target EP attains a higher fidelity, $F_\text{purified}$, if the \texttt{ZZ} operator measures $+1$ and the accrued noise is less than the distilled fidelity gain. Before the application of the second purification protocol, $P_{2}$, the input fidelity of the target EP is overwritten by the value of $F_\text{purified}$. To accomplish this, the idling time in the gate time to create the Bell states with $F_\text{purified}$ is searched over and applied. Then the second purification protocol works with the asymmetric EPs as input. The purification is dynamically applied until the remaining time runs out.

\begin{algorithm}[t] \small
  \caption{Purification protocol selection}
  \label{algo: protocol selection.}
\SetAlgoLined
  \SetKwInOut{Input}{Input}
  \SetKwInOut{Output}{Output}

\BlankLine
  \Input{\texttt{F\textsubscript{in}}: fidelity of raw EP\\
  \texttt{$\tau$}: time consumed for one EP generation\\ 
  \texttt{n}: throughput of one EP generation\\ 
  \texttt{Buffer\_size}: number of qubits in buffer\\ 
  \texttt{Config} \{\texttt{T\textsubscript{1}}, \texttt{T\textsubscript{2}}, \texttt{gate\_length}, \texttt{gate\_err}, ...\}: Configuration of devices\\
  \texttt{T\textsubscript{qos}}: Allotted time to get purified EPs\\ 
  \texttt{F\textsubscript{out}}: Expected fidelity of purified EPs}
  
  \Output{selected protocols $P_\text{out}$ and purified EPs}
  
    $P=\{P_{1},P_{2},P_{3},...\}$ is a list of purification protocols; \\
    $P_\text{out} = [\,]$;
    $T_\text{remain} = $ \texttt{T\textsubscript{qos}};
    Flag = True.\\
\BlankLine    
    \tcc{Check the qualification of protocols}

    \For{$P_{i}$ in $P$}{
        \If{$P_{i}$  has qubits\_needed $>$ \texttt{Buffer\_size}}
        {Trim $P_{i}$ from $P$ }
    }
    \textbf{Return} $P$;
\BlankLine
\tcc{Trim $P$ based on the configuration of the hardware.}

Compute $e_{j}$ from \texttt{Config} \{\texttt{T\textsubscript{1}},\texttt{T\textsubscript{2}},...\}:\\
$e_{j}=\frac{1}{n}\sum_{k=1}^{n}e_{k}$ ($k$: $k$th qubit channel used in $P$)
\BlankLine
\If{$e_{j}>v_{1}$, $\exists j \in J $}{
    Trim $P_{i}$ with more than two rounds
    }
\If{$e_{j}<v_{2}$, $\forall j \in J $}{
    Trim BBPSSW protocols from $P$
    }
\If{$e_{\text{idling}}>v_{3}$, $\forall j \in J$ and $j \neq \text{idling}$}{
    Trim EXPEDIENT protocols from $P$
    } 
\textbf{Return} $P'$;
\BlankLine

\tcc{Sort $P_{i}$ in $P^{'}$}

\While{$T_{\text{remain}} > 0$ and Flag}{
    \eIf{$F_\text{in} \geq F_\text{out}$}{
        break
        }{
        \eIf{$F_{\text{in}} > F_{b}$}{
            Sort DEJMPS protocols to the front of $P'$}
            {Sort EXPEDIENT protocols to the front of $P'$
            }
        \tcc{Run $P_{1}'$}
        \eIf{$F_\text{purified} \leq F_\text{in}$}{
            Flag$=$False, $T_\text{remain} = 0$
            }{
            $F_\text{in} = F_\text{purified}$, $T_\text{remain}=T_\text{remain}-T_{P_{1}'}$,\\
            append $P_{1}'$ to $P_\text{out}$
            }
        }
}
\BlankLine
\textbf{Return} $P_\text{out}$ and EPs
\end{algorithm}

\section{Evaluation}

\subsection{Setup of simulation}

\textbf{a) Simulator}:
The density matrix simulator, NWQ-Sim, built on a classical multi-node, multi-CPU/GPU heterogeneous HPC system is used in this work \cite{li2020density}. It carries a full set of information of quantum states in the simulation and leverages the prompt matrix calculation speed of the HPC system. 

\textbf{b) Noise model}: Our simulation aims to model the purification protocols implementation in a realistic situation. The errors involved can be classified into two categories: 1). The errors of the local operations, local gates, and measurements. 2) The decay and decoherence of the qubits while the qubits are idling. The different types of noises are modeled into quantum channels with Kraus operators shown in Part B of the background section. The values of parameters in Kraus operators are modified utilizing the experimental calibrated parameters of devices encapsulated in a JSON file, including lifetime, \texttt{T\textsubscript{1}}, coherence time, \texttt{T\textsubscript{2}}, the error rate of gates,  \texttt{gate\_err}, the execution time of gates, \texttt{gate\_length}, readout time, \texttt{readout\_length}, measurement errors, \texttt{prob\_meas0\_prep1} and \texttt{prob\_meas1\_prep0}, and coupling map of qubits, \texttt{coupling\_map}. 

Here, the error rate of gates is calibrated from a device as an experimental averaged number. To incorporate it in the simulator, the thermal relaxation error of gates is firstly estimated using \texttt{T\textsubscript{1}}, \texttt{T\textsubscript{2}} and \texttt{gate\_length}. If the thermal relaxation error of a gate is already larger than the experimental calibrated error rate, the depolarizing error is skipped. Otherwise, the rest of the error is deployed as a depolarizing error on gates. 
 
\textbf{c) Benchmark programs}:
The benchmark programs used in this work can be classified into two categories.

The first category consists of three sections of random sampling of all parameters. In the first section, the error parameters of backend devices are randomly sampled from the range that is one order of magnitude smaller than the current state of the art. The second section has the parameters of devices randomly sampled from the ranges of current state of art, that is, the range of the parameters obtained from current devices. The sampling of the third section uses the parameter ranges one order of magnitude larger to verify the performance of our proposed module in a worse situation. Besides the parameters of single devices, the EP generation time $\tau$ is selected from [0, 50] ns, and the throughput of EP generation $n$ is selected from [1, 14] (Details are shown in our repository \cite{githubrepo}). Each section has 100 times run of the protocol selection. The second category of benchmark programs is the configuration of a few backends with the number of qubits larger than 10 from an IBM provider. In the use of these configurations, we assume the coupling map of the backend device is all-to-all connected. 

\textbf{d) Default approach}:
According to previous studies on purification protocols, which either used Clifford gates or assumed the gate error $p$ of \texttt{CX} and measurement error $p_{m}$ of all qubits to be 0.01, the EXPEDIENT protocol has significantly better performance in purifying EPs compared with other types of protocols. Therefore, here, the one round of EXPEDIENT protocol is selected as the Default protocol for comparison with the delivered protocol from our module. Within the allotted time \texttt{T\textsubscript{qos}}, the Default protocol is repeatably implemented as long as it can improve the fidelity of EPs.

\textbf{e) Exhaustive search}:
To demonstrate the advantages of our purification protocol selection module, we use the protocols from the exhaustive search as the upper bound for EP fidelity improvement. In an exhaustive search, the output fidelity of all protocols is exhaustively calculated using the density matrix simulator, and the protocol that offers the best fidelity improvement is selected and applied to raw EPs. Within the allotted time, this exhaustive search approach offers the optimal protocol path to achieve the highest fidelity. If no protocol at any step can improve fidelity, we skip the purification process and retain the original fidelity.

\textbf{f) Metrics}:
This module is designed to pass EPs to meet the expectations of the network layer and facilitate computations on the distributed systems. To characterize the capabilities of our proposed module, we consider the following metrics:

\begin{enumerate}
\item $P_\text{failure}$(\%): The rate of failed selections, where the fidelity of the EP yielded by the selected protocol $P$ is lower than that yielded by the default protocol $P$.
\item $P_\text{success}$(\%): The rate of success selections, where the fidelity of the EP yielded by the selected protocol $P$ is larger than that yielded by the default protocol $P$.
\item $P_\text{nan}$(\%): The rate at which no protocol $P$ can improve the EP where even the exhaustive search fails to deliver protocols capable of enhancing the entanglement due to the limitation of the hardware setup.
\item $P_\text{optimal}$(\%): The rate, determined by the number of instances our proposed module delivered an optimal protocol relative to the total number of successful selections. 
\end{enumerate}

To numerically analyze the output of purification protocol selections, we consider the metrics below:
\begin{enumerate}
\item $\Delta F_\text{max}$: The maximum fidelity improvement among all successful tests;
\item $\Delta F_\text{mean}$: The average fidelity improvement among all tests.
\end{enumerate}

\subsection{Simulation results}
In this section, we analyze the simulation results of our proposed module compared to the Default approach and Exhaustive search approach to demonstrate the advantages of our purification protocols selection module.

\begin{figure}
    \centering
    \includegraphics[width = 3.2 in]{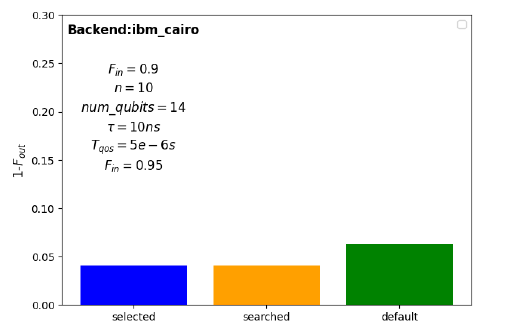}
    \caption{Results of protocol selection based on IBM\_Cairo device properties with all-to-all connectivity. The selected protocol yields an EP with an infidelity smaller than that of the default method and matches that of the exhaustive search.}
    \label{onebackend}
\end{figure}

Figure. \ref{onebackend} shows an example of simulation results using the configuration of IBM backend: ibm\_Cairo. With the assumption of $\texttt{F\textsubscript{in}} = 0.9$, $n = 10$, $\texttt{num\_qubits} = 10$, $\tau = 10 $ ns, $\texttt{T\textsubscript{qos}} = 5\times10^{-6}$ s and $\texttt{F\textsubscript{out}} = 0.95$, both Exhaustive search and our selection module deliver one round of Bennett protocol as best protocol, and yield EP with fidelity 0.96 that over the expectation of \texttt{F\textsubscript{out}}, which beat the performance of the Default protocol. Here one round of Bennett protocol working as the best purification protocol indicates that for a real device with various realistic errors, the implementation of advanced protocols with multiple rounds does not necessarily show better performance, which, contrary to intuition, may damage the entanglement of quantum states.

Table \ref{my_table_1} shows the simulation results of benchmark programs. For the first group of benchmark programs, our purification selection module has rates 69\%, 70\%, and 33\%, respectively, to successfully deliver the protocol with the capability to improve the fidelity of EPs. Meanwhile, it has rates 82\%, 86\%, and 85\%, respectively, for delivering the optimal purification protocol and best quality EPs within the capabilities of the setup among the successful cases. 
The results in the last row of the table show the simulation results using the in-time calibrated parameters of IBMQ real devices. The calibrated configuration of devices is fetched from the IBM provider. The parameters of the buffer and EP generator used are identical to the values used in the example for ibm\_Cairo. Among the tests on five devices, two of the selections deliver the protocol that successfully improves the fidelity of EPs; one of the protocol selections yields the optimal protocol identical to the protocol found by the exhaustive search. The tests on rest two devices show no purification protocol could enhance the fidelity of EPs with their physical constraints. The $P_{nan}$ is 40\% among the test with real device configuration.

\begin{table}[h]
\centering
\begin{tabular}[|c|c|c|c|]{|c|c|c|c|c|c|}
\hline
Type & Group & $P_\text{failure}$ & $P_\text{sucess}$ & $P_\text{nan}$ & $P_\text{opt}$ \\ \hline
\multirow{3}{*}{random sampling} & \#1& 0.04 & 0.69 & 0.27 & 0.82  \\ \cline{2-6}
& \#2  & 0.03 & 0.70 & 0.27 & 0.86  \\ \cline{2-6}
& \#3  & 0.00 & 0.33 & 0.67 & 0.85  \\ \hline
ibmq\_backends & / & 0 & 0.6 & 0.4 & 0.2  \\ \hline
\end{tabular}
\BlankLine
\caption{Simulation results of purification protocol selection module running on the benchmark programs.}
\label{my_table_1}
\end{table}

Table \ref{my_table_2} shows the analytical results obtained from the purified EPs. The maximum fidelity improvements are 15.46\%, 15.43\%, and 8.87\% for tests of three sections of random sampling, respectively, and 5\% for all tested IBM devices. The mean fidelity improvement, denoted as $F_\text{mean}$, is obtained by averaging $\Delta F$ across all cases including the failure and nan cases.

Overall, our purification protocol selection module increases the fidelity of the output entanglement compared with the default method. It has a profound impact on the future development of quantum networks.

\begin{table}[h]
\centering
\begin{tabular}[|c|c|c|c|]{|c|c|c|c|}
\hline
Type & Group & $\Delta F_\text{max}$ & $\Delta F_\text{mean}$   \\ \hline
\multirow{3}{*}{random sampling} & \#1 & 0.1507 & 0.0520 \\ \cline{2-4}
& \#2  & 0.1543 & 0.0522  \\ \cline{2-4}
& \#3  & 0.0887 & 0.0066  \\ \hline
ibmq\_backends & /  & 0.0589
 & 0.0284
  \\ \hline
\end{tabular}
\BlankLine
\caption{Analytical results of purification protocol selection module running on the benchmark programs.}
\label{my_table_2}
\end{table}

\section{Discussion}

To the best of our knowledge, this is the first work that considers the realistic errors of devices, the size of quantum storage, and EP generation mechanisms in the entanglement purification protocol selection. This protocol selection avoids exhaustive tests of a multitude of purification protocols but directly selects the appropriate protocols for a hardware setup. Usage of the protocol selection module significantly improves inter-node communications quality and therefore facilitates practical applications on quantum networks. We believe this paper as a heuristic module would lead the theoretical studies of entanglement purification to play an important role in distributed quantum computation and quantum network applications.

Although we have demonstrated the benefits of our proposed module. There is still space left for future works to explore.

\textbf{a) Expanding the spectrum of applications}.
This work mainly focuses on developing a purification protocol selection module for entanglement shared by two parties, assuming the performance of local operations on two parties is identical. Future work should consider more complex scenarios where entangled states are shared among multiple parties characterized by disparate performances potentially caused by the different types of nodes. Thus, future work should move beyond viewing errors in an aggregate, including the performance discrepancies between parties, which have a deterministic effect on the protocol selection.

\textbf{b) Integration of diverse errors}.
Our selection module accommodates a spectrum of physical layer errors, such as depolarizing, phase damping, and amplitude damping errors, alongside measurement inaccuracies. However, as quantum computers evolve, it becomes imperative to consider a broader array of error types. These include leakage errors from the change of energy level structure in superconducting qubits, cross-talk errors when operations on one qubit unintentionally affect another qubit, and spontaneous emission referred to the process by which an excited qubit loses energy by emitting a photon. Considering a broader range of error types would undoubtedly improve the performance of the purification protocol selection.

\textbf{c) Advancements towards fault-tolerant devices}.
This work focuses on basic devices that do not incorporate quantum error correction codes. Future research should extend the scope of the purification protocol selection to include fault-tolerant devices by exploring methods for their integration. The inclusion of error correction codes will significantly alter the design principles of the purification protocol selection module.

\section{Conclusion}
In this work, we conduct a systematic study of the performance of various purification protocols under realistic errors using the density matrix simulation. Based on the understanding of simulation results, we develop an entanglement purification protocol selection module that considers not only buffer size, EP generator, but also the configuration of the processor to select the potential optimal purification protocols for enhancing the quality of EPs for the quantum network. We verify our proposed module from two types of benchmark programs. The results present a high success rate of our proposed module for selecting the optimal protocols for various hardware setups.

\section{Acknowledgement}
This material is based upon work supported by the U.S. Department of Energy, Office of Science, National Quantum Information Science Research Centers, Co-design Center for Quantum Advantage (C2QA) under contract number DE-SC0012704, (Basic Energy Sciences, PNNL FWP 76274). This research used resources of the National Energy Research Scientific Computing Center (NERSC), a U.S. Department of Energy Office of Science User Facility located at Lawrence Berkeley National Laboratory. This research used resources of the Oak Ridge Leadership Computing Facility, which is a DOE Office of Science User Facility supported under Contract DE-AC05-00OR22725. The Pacific Northwest National Laboratory is operated by Battelle for the U.S. Department of Energy under Contract DE-AC05-76RL01830.

\bibliographystyle{plain}
\bibliography{mybib}

\onecolumn\newpage

\end{document}